\documentclass[prd,
a4paper,
nofootinbib,
preprintnumbers,
twocolumn,
]{revtex4}

\usepackage{amsmath}
\usepackage{amssymb}
\usepackage{epsfig}
\usepackage{graphicx}
\usepackage{mathrsfs}

\newcommand{\dma}{\Delta m_{31}^2}
\newcommand{\dms}{\Delta m_{21}^2}

\newcommand{\eVq}  {\text{eV}^2}

\begin{document}
\title{Global fits to neutrino oscillation data\footnote{Talk 
presented at SNOW2006, Stockholm, 2--6 May 2006}}
\author{Thomas Schwetz}\email{schwetz@sissa.it}
\affiliation{Scuola Internazionale Superiore di Studi Avanzati,
via Beirut 2--4, 34014 Trieste, Italy}
%
\begin{abstract}
   I summarize the determination of neutrino oscillation parameters
   within the three-flavor framework from world neutrino oscillation
   data with date of May 2006, including the first results from the
   MINOS long-baseline experiment. It is illustrated how the
   determination of the leading ``solar'' and ``atmospheric''
   parameters, as well as the bound on $\theta_{13}$ emerge from an
   interplay of various complementary data sets. Furthermore, I
   discuss possible implications of sub-leading three-flavor effects
   in present atmospheric neutrino data induced by $\Delta m^2_{21}$
   and $\theta_{13}$ for the bound on $\theta_{13}$ and non-maximal
   values of $\theta_{23}$, emphasizing, however, that these effects
   are not statistically significant at present. Finally, in view of
   the upcoming MiniBooNE results I briefly comment on the problem to
   reconcile the LSND signal.
\end{abstract}
\maketitle

\section{Introduction}

In the last ten years or so we have witnessed huge progress in neutrino
oscillation physics. The outstanding experimental results lead to
quite a clear overall picture of the neutrino sector. We know that
there are two mass-squared differences separated roughly by a factor
of 30, and in the lepton mixing matrix there are two large mixing
angles, and one mixing angle which has to be small.
In this talk I review the present status of neutrino oscillations by
reporting the results of a global analysis of latest world neutrino
oscillation data from solar~\cite{solar,ahmad:2002ka,sno2005},
atmospheric~\cite{atm,Ashie:2005ik}, reactor~\cite{kamland,chooz}, and
accelerator~\cite{k2k, minos} experiments, including the recent data
from the MINOS long-baseline experiment~\cite{minos}. This analysis is
performed in the three-flavor framework and is based on the work
published in Refs.~\cite{Maltoni:2003da,Maltoni:2004ei} (see also the
hep-ph archive version~5 of Ref.~\cite{Maltoni:2004ei} for updated
results). 
In Sec.~\ref{sec:leading} I discuss the determination of the leading
``solar'' and ``atmospheric'' parameters, whereas Sec.~\ref{sec:th13}
deals with the bound on $\theta_{13}$ from global data. The status of
three-flavor oscillation parameters is summarized in
Tab.~\ref{tab:3nu-summary}. In Sec.~\ref{sec:subleading} a discussion
of sub-leading effects in atmospheric data is given, and in
Sec.~\ref{sec:lsnd} I comment on attempts to reconcile the result of
the LSND experiment~\cite{Aguilar:2001ty} with the global oscillation
data.

\section{Leading oscillation parameters}
\label{sec:leading}

\begin{figure*}
\centering
\includegraphics[height=5.1cm]{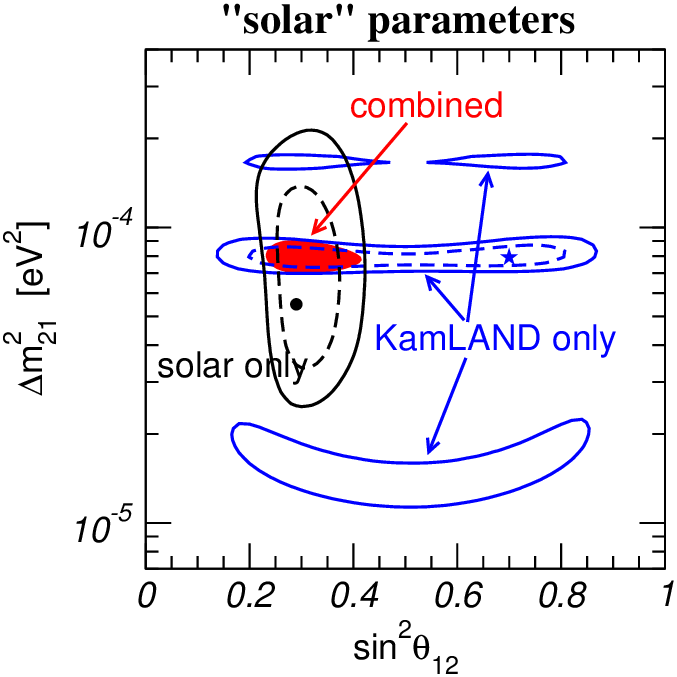}
\quad
\includegraphics[height=5.1cm]{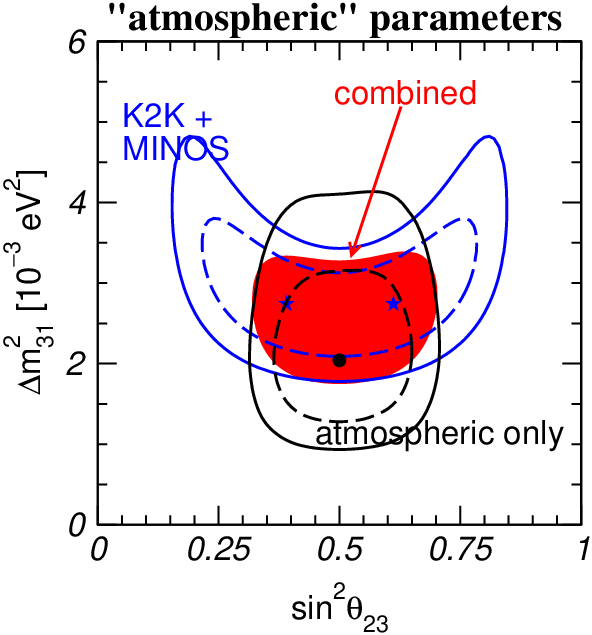}
\quad
\includegraphics[height=5.1cm]{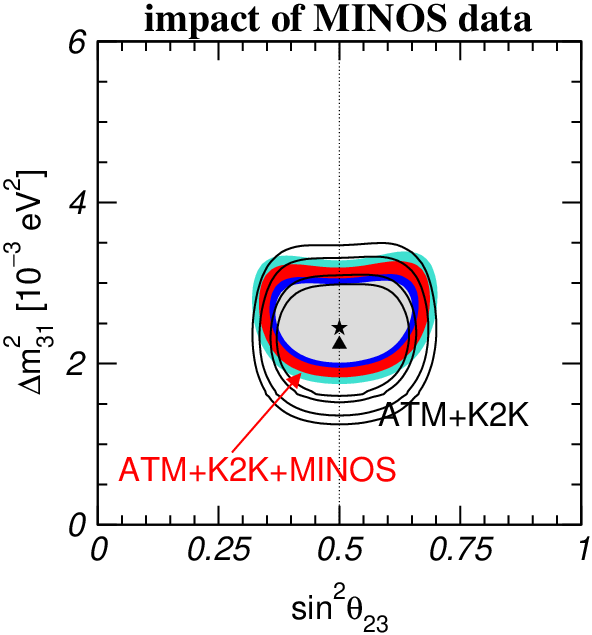}
  \caption{Determination of the leading oscillation parameters from an
  interplay of experiments with natural and artificial neutrino
  sources (left and middle panels). In the right panel the allowed
  regions are shown with (colored regions) and without (contour
  curves) MINOS data. In the left and middle panels the allowed
  regions are shown at 90\%~CL (dashed curves) and 99.73\%~CL (solid
  curves and shaded regions), whereas in the right panel regions are
  shown at 90\%, 95\%, 99\%, and 99.73\%~CL.} \label{fig:leading}
\end{figure*}

In this section I discuss the determination of the leading oscillation
parameters, the ``solar'' parameters $\theta_{12}$, $\dms$, and the
``atmospheric'' parameters $\theta_{23}$, $\dma$. In both cases we
have an independent confirmation of neutrino oscillations from very
different experiments, and the final allowed regions for the
oscillation parameters emerge from an interplay of complementary
data: The determination of the mixing angle is dominated by
experiments with natural neutrino sources (solar and atmospheric
neutrinos), whereas the mass-squared differences are more accurately
determined by man-made neutrinos (from reactors and accelerators).
This complementarity is illustrated in Fig.~\ref{fig:leading}.

Details of our solar neutrino analysis can be found in
Ref.~\cite{Maltoni:2003da} and references therein. We use data from
the Homestake, SAGE, GNO, and SK experiments~\cite{solar}, and
the SNO day-night spectra from the pure D$_2$O
phase~\cite{ahmad:2002ka}, but the CC, NC, and ES rates from the SNO
salt-phase are updated according to the latest 2005
data~\cite{sno2005}. The predictions for the solar neutrino
fluxes are taken from Ref.~\cite{Bahcall:2004pz}.
For the KamLAND analysis we are using the data from
Ref.~\cite{kamland} equally binned in $1/E_\mathrm{pr}$
($E_\mathrm{pr}$ is the prompt energy deposited by the positron), and
we include earth matter effects and flux uncertainties following
Ref.~\cite{Huber:2004xh} (see the appendix of
Ref.~\cite{Maltoni:2004ei} for further details). 

We observe from Fig.~\ref{fig:leading} (left) a beautiful agreement of
solar and KamLAND data. Moreover, the complementarity of the two data
sets allows a rather precise determination of the oscillation
parameters: The evidence of spectral distortion in KamLAND data
provides a strong constraint on $\Delta m^2_{21}$, and leads to the
remarkable precision of $4\%$ at $1\sigma$ (compare
Tab.~\ref{tab:3nu-summary}). Alternative solutions around $\Delta
m^2_{21}\sim 2\times 10^{-4}$~eV$^2$ ($\sim 1.4\times
10^{-5}$~eV$^2$), which are still present in the KamLAND-only analysis
at 99\% C.L., are ruled out from the combined KamLAND+solar analysis
at about $4\sigma$ ($5\sigma$). In contrast to $\Delta m^2_{21}$, the
determination of the mixing angle is dominated by solar data.
Especially recent results from the SNO experiment provide a strong
upper bound on $\sin^2\theta_{12}$, excluding maximal mixing at more
than $5\sigma$.

\begin{table*}
\centering
\begin{tabular}{@{\quad}l@{\qquad}c@{\qquad}c@{\qquad}c@{\qquad}c@{\quad}}
  \hline\hline
  parameter & bf$\pm 1\sigma$ & $1\sigma$ acc. & 2$\sigma$ range & 3$\sigma$ range\\
  \hline
  $\Delta m^2_{21}  \: [10^{-5}\eVq]$ & $7.9\pm 0.3$          &  4\% 
     & $7.3-8.5$ & $7.1-8.9$ \\
  $|\Delta m^2_{31}|\: [10^{-3}\eVq]$ & $2.5^{+0.20}_{-0.25}$ & 10\%
     & $2.1-3.0$ & $1.9-3.2$ \\
  \hline
  $\sin^2\theta_{12}$ & $0.30^{+0.02}_{-0.03}$ &  9\% & $0.26-0.36$ & $0.24-0.40$ \\
  $\sin^2\theta_{23}$ & $0.50^{+0.08}_{-0.07}$ & 16\% & $0.38-0.64$ & $0.34-0.68$ \\
  $\sin^2\theta_{13}$ & $-$  &$-$  & $\leq 0.025$ & $\leq 0.041$ \\
  \hline\hline
\end{tabular}
\caption{Best fit values (bf), $1\sigma$ errors, relative accuracies
  at $1\sigma$, and $2\sigma$ and $3\sigma$ allowed ranges of
  three-flavor neutrino oscillation parameters from a combined
  analysis of global data.} \label{tab:3nu-summary}
\end{table*}

Oscillations with the ``atmospheric'' parameters $\theta_{23}$ and
$\dma$ have been established by the atmospheric neutrino data of
SK~\cite{atm}. Details of our re-analysis of the SK-I zenith angle
distributed data can be found in Ref.~\cite{Maltoni:2004ei} and
references therein. The allowed region is shown in
Fig.~\ref{fig:leading} (middle). Also In this case, by now we have an
independent confirmation of the effect by experiments based on
man-made neutrinos, namely the first generation of long-baseline (LBL)
accelerator experiments exploring the $\nu_\mu$ disappearance
oscillation channel.
In the K2K experiment~\cite{k2k} the neutrino beam is produced at the
KEK proton synchrotron, and originally consists of 98\% muon neutrinos
with a mean energy of 1.3~GeV. The $\nu_\mu$ content of the beam is
observed at the SK detector at a distance of 250~km. For the
K2K-I and K2K-II data ($0.89\times 10^{20}$~p.o.t.\ in total) 107
events have been detected, whereas $151^{+12}_{-10}$ have been
expected for no oscillations. 

Recently first data ($0.93\times 10^{20}$~p.o.t.) from the MINOS
experiment have been released~\cite{minos}. A neutrino beam with
$98.5\%~(\nu_\mu + \bar\nu_\mu)$ and a mean energy of 3~GeV is
produced at Fermilab and observed at the MINOS detector in the Soudan
mine at a distance of 735~km. In the absence of oscillations
$177\pm11$ $\nu_\mu$ events with $E<10$~GeV are expected, whereas 92
have been observed, which provides a $5.0\sigma$ evidence for
disappearance. In our re-analysis we use spectral data divided into 15
bins in reconstructed neutrino energy, and our allowed region from
MINOS-only is in very good agreement with the official
result~\cite{minos}. The values of the oscillation parameters from
MINOS are consistent with the ones from K2K, as well as from SK
atmospheric data. The impact of the data from MINOS in the global
analysis is illustrated in Fig.~\ref{fig:leading} (right). We find
that the best fit point for $\dma$ is shifted upward from $2.2\times
10^{-3}$~eV$^2$ for SK+K2K to $2.5\times 10^{-3}$~eV$^2$. In addition
MINOS improves the lower bound on $\dma$, which is increased from
$1.4\times 10^{-3}$~eV$^2$ for SK+K2K to $1.9\times 10^{-3}$~eV$^2$ at
$3\sigma$. The relative accuracy on $\dma$ at $1\sigma$ is improved
from 14\% to 10\%. As obvious from the middle panel of
Fig.~\ref{fig:leading} the determination of $\theta_{23}$ is
completely dominated by atmospheric data and there is no change due to
MINOS. Let us add that present data cannot distinguish between $\dma >
0$ and $<0$, and hence, both, the normal and inverted neutrino mass
hierarchies provide equally good fits to the data.

\section{The bound on $\theta_{13}$}
\label{sec:th13}

\begin{figure}
\centering
\includegraphics[width=0.38\textwidth]{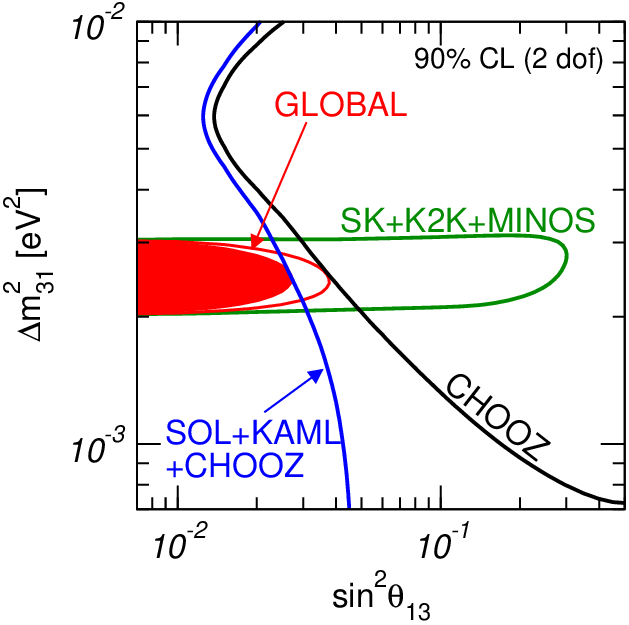}
  \caption{The bound on $\sin^2\theta_{13}$ from the 
  interplay of the global data.} 
  \label{fig:th13}
\end{figure}

Similar to the case of the leading oscillation parameters, also the
bound on $\theta_{13}$ emerges from an interplay of different data
sets, as we illustrate in Fig.~\ref{fig:th13}. An important
contribution to the bound comes, of course, from the CHOOZ reactor
experiment combined with the determination of $\dma$ from atmospheric
and LBL experiments. However, due to a complementarity of low and high
energy solar data, as well as solar and KamLAND data also
solar+KamLAND provide a non-trivial constraint on $\theta_{13}$, see
e.g., Refs.~\cite{Maltoni:2003da,Maltoni:2004ei,Goswami:2004cn}. We
find at 90\%~CL ($3\sigma$) the following limits:
%
\[
  \sin^2\theta_{13} < \left\{
  \begin{array}{l@{\qquad}l}
     0.027 \:(0.058) & \mbox{CHOOZ+atm+LBL,}\\
     0.033 \:(0.071) & \mbox{solar+KamLAND,}\\
     0.020 \:(0.041) & \mbox{global data.}  
  \end{array}
  \right.
\]
The addition of MINOS data leads to a slight tightening of the
constraint (the $3\sigma$ limit from CHOOZ+atm+K2K is shifted from
0.067 to 0.058 if MINOS is added) because of the stronger lower bound
on $\dma$, where the CHOOZ bound becomes weaker (c.f.\
Fig.~\ref{fig:th13}). Note that also the update in the solar
model~\cite{Bahcall:2004pz} leads to a small shift in the limit from
solar+KamLAND data (from 0.079 to 0.071 at $3\sigma$). Both of these
updates contribute to the change of the global bound from
0.046~\cite{Schwetz:2005jr} to 0.041 at $3\sigma$.

\section{Sub-leading effects in atmospheric neutrinos}
\label{sec:subleading}

In principle one expects that at some level sub-leading effects will
show up in atmospheric neutrinos, involving oscillations with $\dms$
or effects of a finite $\theta_{13}$, see e.g.,
Refs.~\cite{atm13,atm-solar-osc,Gonzalez-Garcia:2004cu,Fogli:2005cq,kajita}.
An excess of $e$-like events observed in
SK~\cite{Ashie:2005ik} might be a possible hint for such effects, and
in Refs.~\cite{Gonzalez-Garcia:2004cu,Fogli:2005cq} a slight
preference for non-maximal values of $\theta_{23} < \pi/4$ has been
found. In contrast, the SK analysis presented in Ref.~\cite{kajita}
did not confirm that hint.

\begin{figure}
\centering
\includegraphics[width=0.45\textwidth]{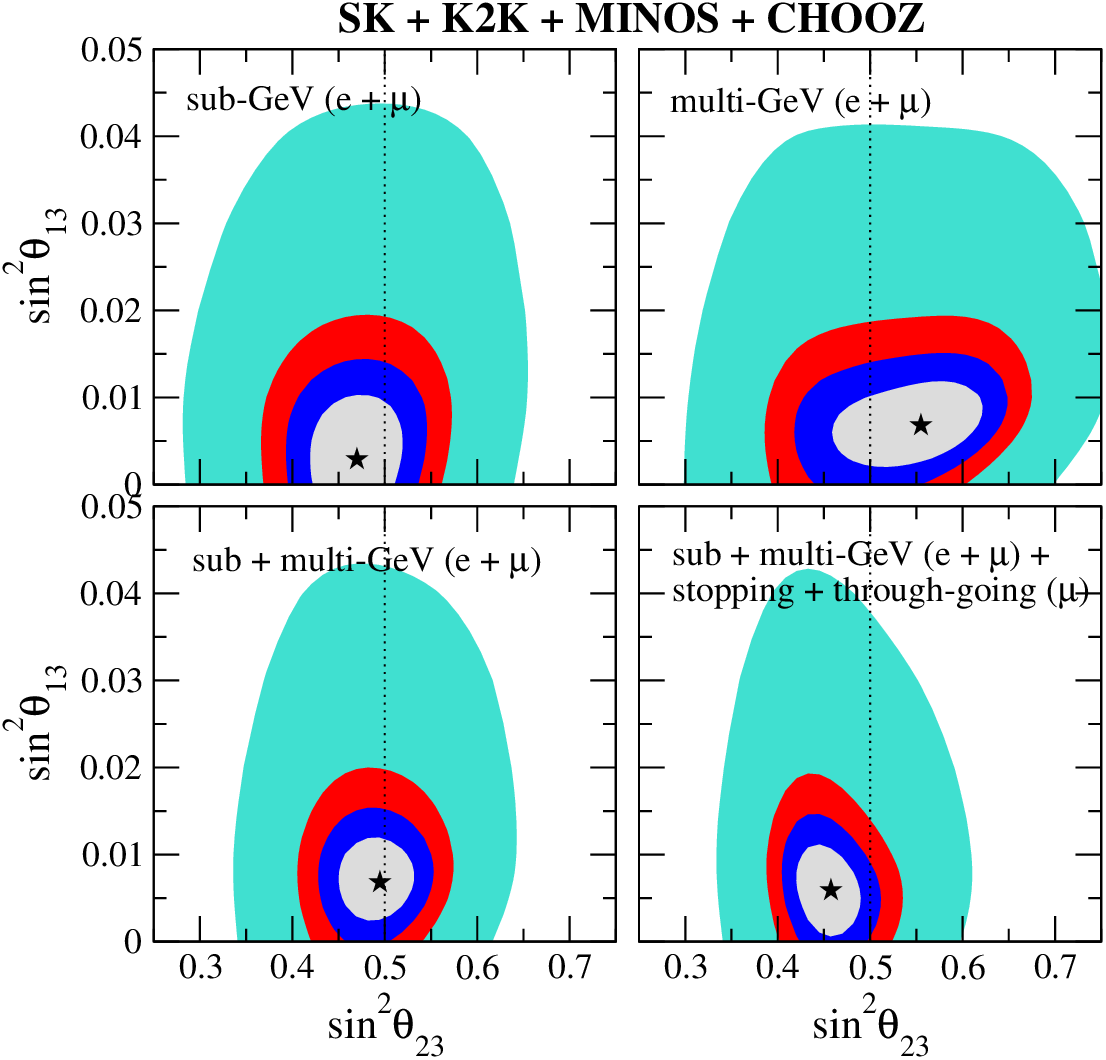}
  \caption{Contours of $\Delta\chi^2 = 0.2, 0.5, 1, 4.6$ in the plane 
  $\sin^2\theta_{23}$-$\sin^2\theta_{13}$ from various SK data
  samples, taking into account oscillations with $\dms = 8\times
  10^{-5}$~eV$^2$.} \label{fig:th13-th23}
\end{figure}

From a full three-flavor analysis of SK data~\cite{michele} shown in
Fig.~\ref{fig:th13-th23} one finds that indeed sub-GeV data prefer a
value $\theta_{23} < \pi/4$, however, if only multi-GeV data is used
the best fit occurs for $\theta_{23} > \pi/4$. Summing sub- and
multi-GeV data leads incidentally to a cancellation of both effects
and the best fit occurs very close to maximal mixing. Finally, using
all data including sub-GeV, multi-GeV, stopping and through-going
$\mu$-like data, the best fit moves again to $\sin^2\theta_{23} =
0.46$~\cite{Gonzalez-Garcia:2004cu}. From these considerations we
conclude that the final result for $\theta_{23}$ appears as a delicate
interplay of different data samples, involving cancellations of
opposite trends. Hence the result is rather sensitive to the very fine
details of the analysis. Let us stress that the $\Delta\chi^2$
contours shown in Fig.~\ref{fig:th13-th23} correspond to 9.5\%, 22\%,
39\%, and 90\%~CL (2 d.o.f.), i.e., there is no significance in these
effects. The purpose of this analysis is to show that present data
does not allow to obtain statistically meaningful indications of
non-maximal values of $\theta_{23}$ nor of non-zero values of
$\theta_{13}$. Nevertheless, sub-leading three-flavor effects in
atmospheric oscillations can be explored in future Mt scale water
\v{C}erenkov~\cite{Huber:2005ep} or magnetized iron
calorimeter~\cite{Petcov:2005rv} experiments, and may provide
complementary information to LBL experiments.

\begin{figure}
\centering
\includegraphics[width=0.47\textwidth]{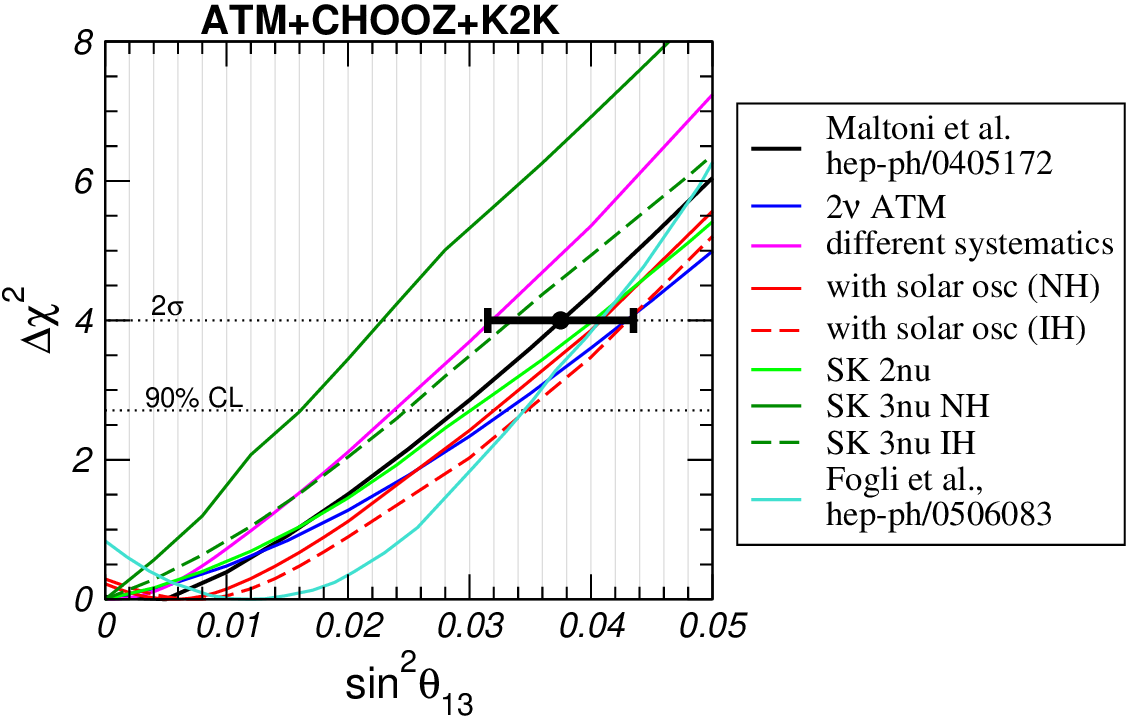}
  \caption{Dependence of the bound on $\sin^2\theta_{13}$ from
  CHOOZ+atm+K2K data on the details of the atmospheric neutrino
  analysis. The curves (from top to bottom in the legend) correspond
  to {\it (i)} the analysis of Ref.~\cite{Maltoni:2004ei} without
  $\dms$-effects, {\it (ii)} a 2-flavor analysis neglecting
  $\theta_{13}$ effects in ATM, {\it (iii)} an analysis with a
  different systematical error
  treatment~\cite{Gonzalez-Garcia:2004wg}, {\it (iv)} taking into
  account the effect of $\dms$ for both types of the neutrino mass
  hierarchy, {\it (v)} using the 2-flavor SK
  analysis~\cite{Ashie:2005ik}, {\it (vi)} the SK analysis~\cite{sk3nu}
  taking into account $\theta_{13}$ effects for both hierarchies, and
  {\it (vii)} the analysis from Ref.~\cite{Fogli:2005cq}.}
  \label{fig:th13-atm}
\end{figure}

Fig.~\ref{fig:th13-atm} illustrates how details of the atmospheric
neutrino analysis affect the bound on $\sin^2\theta_{13}$ from
CHOOZ+atm+K2K data. It is evident from the figure that the inclusion
of three-flavor effects (from $\theta_{13}$ and/or $\dms$), as well
as different treatments of systematics lead to an
``uncertainty'' of about 16\% on the bound on $\sin^2\theta_{13}$ at
$2\sigma$, as indicated by the ``error bar'' in the figure. Note that
the shifts of the global $\theta_{13}$ limit due to MINOS or changes
in the solar neutrino analysis reported in Sec.~\ref{sec:th13} are at
the same level as this uncertainty from details in the atmospheric
neutrino analysis.

\section{The LSND problem}
\label{sec:lsnd}

To reconcile the LSND evidence~\cite{Aguilar:2001ty} for
$\bar\nu_\mu\to\bar\nu_e$ oscillations with $\Delta m^2 \sim$~eV$^2$
is a long-standing problem for neutrino phenomenology, and the
community is eagerly waiting for an experimental answer to this
problem from the MiniBooNE experiment~\cite{miniboone}. The three
required mass-squared differences can be obtained in four-neutrino
mass schemes, but such models cannot accommodate the constraints on
the mixing~\cite{Maltoni:2002xd} (see Ref.~\cite{Maltoni:2004ei} for
an updated analysis): Mass schemes of the (2+2) type predict that a
large fraction of the sterile neutrino participates in solar and/or
atmospheric neutrino oscillations, which in both cases is disfavored
by the data~\cite{Maltoni:2002xd,Gonzalez-Garcia:2001uy}, and
therefore such schemes are ruled out at more than $5\sigma$~CL. The
(3+1) mass spectra are in perfect agreement with solar and atmospheric
data, however, they suffer from a tension between the LSND signal and
null-result short-baseline disappearance
experiments~\cite{3+1,Bilenky:1999ny}, most importantly
Bugey~\cite{Declais:1994su} and CDHS~\cite{Dydak:1983zq}, which
disfavors these models at the $3\sigma$ level.

\begin{figure}
\centering
\includegraphics[width=0.35\textwidth]{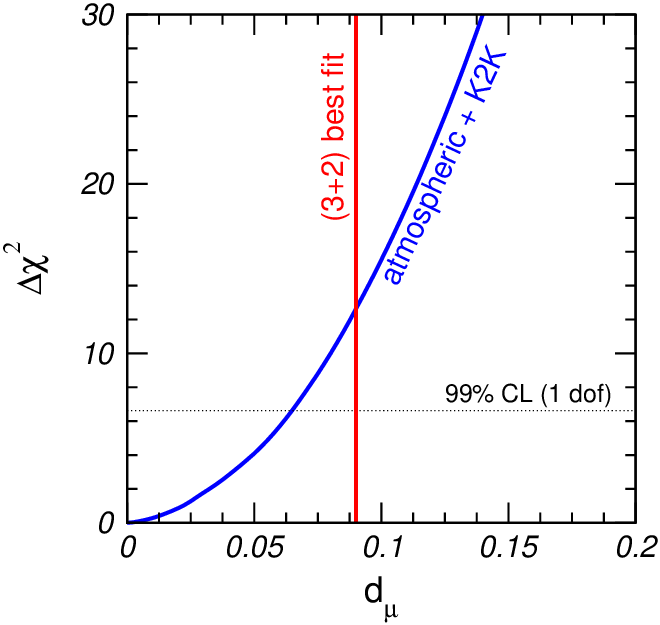}
  \caption{Constraint on the parameter $d_\mu$ from atm+K2K data from
  Ref.~\cite{Maltoni:2004ei} compared to the best fit prediction of
  the (3+2) neutrino mass scheme~\cite{Sorel:2003hf}.}\label{fig:d_mu}
\end{figure}

In Ref.~\cite{Sorel:2003hf} a five-neutrino mass scheme of the type
(3+2) has been considered to avoid these constraints, and it is
claimed that the disagreement measured by the so-called parameter
goodness-of-fit~\cite{Maltoni:2003cu} is improved from 0.032\% for
(3+1) to 2.1\%. However, it should be noted that, apart from possible
severe conflicts with constraints from cosmology, the best fit point
found in Ref.~\cite{Sorel:2003hf} seems to be disfavored also from
atmospheric neutrino data. As pointed out in
Ref.~\cite{Bilenky:1999ny} atmospheric neutrinos provide a constraint
on a parameter $d_\mu$, denoting the fraction of $\nu_\mu$ which does
not participate in oscillations with $\Delta m^2_\mathrm{atm}$. In the
(3+2) scheme this parameter is given by $d_\mu = |U_{\mu 4}|^2 +
|U_{\mu 5}|^2$, and with the best fit values~\cite{Sorel:2003hf}
$U_{\mu 4} = 0.204$, $U_{\mu 5} = 0.224$ one finds $d_\mu \approx
0.09$. As visible from Fig.~\ref{fig:d_mu} this value leads to a
$\Delta\chi^2 \approx 12.5$ from atmospheric+K2K data, and hence seems
to be disfavored at the $3.5\sigma$ level. Therefore, a re-analysis
of the (3+2) scenario including the constraint from atmospheric data
seems to be required to judge the viability of this model.

In view of these difficulties to explain the LSND result with neutrino
oscillations several alternative mechanisms have been proposed, see
Ref.~\cite{Palomares-Ruiz:2005vf} for references. In addition to the
fact that some of them involve very speculative physics, many of these
proposals have also phenomenological problems to accommodate all
constraints. Scenarios which seem to be in agreement with all present
data are a model with a decaying sterile
neutrino~\cite{Palomares-Ruiz:2005vf}, a four-neutrino mass scheme
plus CPT violation~\cite{4nu-cpt}, and a model based on sterile
neutrinos and large extra dimensions~\cite{Pas:2005rb}.

\section{Summary}

In this talk I have summarized the status of neutrino oscillations
with May 2006, providing updated best fit values and allowed ranges of
the three-flavor neutrino oscillation parameters. The impact of the
recently released first data from MINOS on the determination of $\dma$
as well as on the bound on $\theta_{13}$ has been
investigated. Furthermore, sub-leading effects in atmospheric neutrino
data have been discussed, stressing that hints for non-maximal values
of $\theta_{23}$ and/or non-zero values of $\theta_{13}$ depend on the
fine details of the analysis and are not statistically significant.
In view of the upcoming results from MiniBooNE I have reviewed once
again the problem related to the LSND result, and a confirmation of
the effect by MiniBooNE would imply a serious challenge to neutrino
oscillation phenomenology.

{\bf Acknowledgments.} 
I thank the organizers for the very pleasant and stimulating
workshop. The results presented here have been obtained in
collaboration with M.~Maltoni, M.A.~T{\'o}rtola, and J.W.F.~Valle, and
in particular I would like to thank M.~Maltoni for the permission to
use his atmospheric neutrino code for the analysis of sub-leading
effects. T.S.\ is supported by a ``Marie Curie Intra-European
Fellowship within the 6th European Community Framework Program.''

\end{document}